\documentclass[twocolumn,english,aps,prl,groupedaddress,superscriptaddress]{revtex4}
\usepackage[pdftex]{graphicx} %include PDF or jpg figures
\usepackage{epstopdf}
\usepackage{xcolor} %colored text, {\color{blue}......} 
\usepackage{soul} %stikeout lines, \st{text}
\usepackage{amsmath,amsfonts,mathrsfs,amsbsy,bm,babel}
\usepackage{dcolumn}% Align table columns on decimal point
\usepackage{bm}% bold math
\usepackage[utf8x]{inputenc}

\begin{document}
\title{Spin reorientation and Ce-Mn coupling in antiferromagnetic oxypnictide 
CeMnAsO}
\author{Qiang Zhang}\email{qzhangemail@gmail.com} 
\affiliation{Ames Laboratory, Ames, IA, 50011, USA}
\affiliation{Department of Physics and Astronomy, Iowa State University, Ames, IA, 50011, USA}

\author{Wei Tian}
\affiliation{Oak Ridge National Laboratory,Oak Ridge,Tennessee 37831, USA}

\author{Spencer G. Peterson}
\affiliation{Ames Laboratory, Ames, IA, 50011, USA}
\affiliation{Department of Physics and Astronomy, Iowa State University, Ames, 
IA, 50011, USA}
\author{Kevin W. Dennis}
\affiliation{Ames Laboratory, Ames, IA, 50011, USA}
\affiliation{Department of materials and engineering, Iowa State University, Ames, IA, 50011, USA}

\author{David Vaknin}\email{vaknin@ameslab.gov}
\affiliation{Ames Laboratory, Ames, IA, 50011, USA}
\affiliation{Department of Physics and Astronomy, Iowa State University, Ames, IA, 50011, USA}

\date{\today}

\begin{abstract}
Structure and magnetic properties of high-quality polycrystlline CeMnAsO, a 
parent compound  of the  ``1111"-type oxypnictides, have been  
investigated using neutron  powder diffraction and magnetization measurements. 
We find that  CeMnAsO undergoes a C-type antiferromagnetic order with Mn$^{2+}$ 
($S=5/2$) moments pointing along the \textit{c}-axis below a relatively high 
N\'{e}el temperature of $T_{\rm N} = 347(1)$ K.  Below $T_{\rm SR} = 35$ K, two 
simultaneous transitions occur where the Mn moments reorient from the 
$c$-axis to the \textit{ab}-plane preserving the C-type magnetic order, and Ce 
moments undergo long-range AFM ordering with antiparallel moments pointing in 
the \textit{ab}-plane.  Another transition to a noncollinear magnetic 
structure occurs below  7 K. The ordered moments of Mn and Ce at 2 K are 3.32(4) $\mu_{B}$ 
and 0.81(4)$\mu_{B}$,  respectively.  We find that CeMnAsO primarily falls into 
the category of a local-moment antiferromagnetic insulator in which the 
nearest-neighbor interaction  ($J_{1}$) is dominant with 
$J_{2}<J_{1}/2$ in the context of $J_{1}-J_{2}-J_{c}$ model. The 
spin reorientation transition driven by the coupling between Ce and 
the transition metal seems to be common to Mn, Fe and Cr ions, but not to Co and Ni 
ions in the iso-structural oxypnictides.  A schematic illustration of magnetic structures in Mn and Ce 
sublattices in CeMnAsO is presented.
 
\end{abstract}
 \maketitle
\section{INTRODUCTION}
The discovery of superconductivity (SC) in LaFeAsO$_{1-y}$F$_{y}$\cite{Kamihara2008}    has  triggered renewed interest in  
superconductivity and  also in itinerant magnetism  in general. In the Fe-based 
``1111'' and ``122" pnictides, the emergence of 
superconductivity is accompanied by the suppression of the stripe-like 
antiferromagnetic (AFM) ordering of Fe$^{2+}$  and  a 
tetragonal(T)-orthorhombic(O) structural transition\cite{review4}. It has  also 
been suggested that AFM/structural fluctuations may be the driving forces for 
superconductivity\cite{reviews_pairing}. In view of the prominent role of 
magnetism in driving superconductivity in the Fe-based pnictides, in particular 
by doping with transition metal ions, systematic investigations of the magnetic 
and structural properties of the  iso-structural ``1111'' and ``122'' parent pnictides  involving the square lattice of other transition metals are called for\cite{Zhao2009}. In
fact, the various transition metals  influence subtle and specific structural, 
magnetic, and electronic properties, which provides insight to 
understanding how the 
magnetism is related to the SC state\cite{Ohta2009}. Furthermore, compared with 
the parent ``122" compounds, the parent ``1111" oxypnictides \textit{RT}AsO  
(\textit{R} = magnetic and non-magnetic rare earths, $T =$ transition metals Fe, 
Co, Ni, Mn, etc.) are more intriguing  as they offer the possibility to tweak an 
additional coupling between the rare-earth \textit{R} and the 
transition-metal ions. 
 
CeFeAsO  is an itinerant poor metal with a T-O structural transition at $T_S \approx 
150$ K followed by stripe-like AFM order of Fe at $T_N \approx 145$ K\cite{Zhao2008}. 
Recent $\mu SR$ studies\cite{Maeter2009} have indicated relatively strong 
coupling between the rare earth Ce and Fe, and further neutron and X-ray 
scattering studies have shown the coupling leads to a gradual Fe 
spin-reorientation at low temperatures\cite{Zhang2013}. By contrast, in the 
heavy-fermion metal CeNiAsO\cite{Luo2014}, Ni does not order magnetically but 
two successive AFM transitions associated with Ce ions are 
observed\cite{Luo2011}.  In CeCoAsO, a ferromagnetic ordering is found below 
$\sim75$ K with no indication for Ce ordering at lower 
temperatures\cite{Sarkar2010}. To date, little attention has been paid to 
CeMnAsO for which magnetization and heat capacity measurements indicate that  Mn 
moments order above room temperature and a first-order magnetic transition 
emerges at $\sim$  35 K, possibly related to a Mn spin 
reorientation\cite{TSUKAMOTO2011}. It has also been proposed that the Ce spins 
do not undergo long range order but are {\it parasitically induced to order} below  
$\sim$ 35 K\cite{TSUKAMOTO2011}.  However, the AFM N\'{e}el temperature $T_N$, 
actual magnetic structures, the values of the ordered moments, and  
the interplay between Ce and Mn in CeMnAsO have not been determined. Here, we 
report neutron diffraction, and magnetization results on CeMnAsO to answer these 
questions, and also to critically compare the structure and magnetism with 
related pnictides.
\section{EXPERIMENTAL DETAILS}
Previous reports on the synthesis of CeMnAsO used 
CeAs and Mn$_{2}$O$_{3}$ \cite{TSUKAMOTO2011} and added excess Ti as an oxygen 
getter which resulted in the formation of CeMnAsO with a secondary phase. In  
the present 
study,  MnO and CeAs as starting materials were mixed thoroughly in 
stoichiometric proportions. (CeAs was prepared firstly by reacting Ce and As 
powders at 600 $^{\rm o}$C for 35 h and then at 950 $^{\rm o}$C for 8 h). The mixed powder was 
sealed in an evacuated tantalum tube and sintered at 1150 $^{\rm o}$C for 40 h. 
A single-phase polycrystalline CeMnAsO powder was then obtained and 
characterized by x-ray and neutron diffraction methods.

Neutron powder diffraction (NPD) measurements on  $\approx 4$ g CeMnAsO sample 
were conducted on the HB1A  triple-axis spectrometer with a fixed-incident-energy 
14.6 meV (located at the high flux isotope reactor, HFIR, at the Oak Ridge 
National Laboratory, USA). The measurements on HB1A were performed with an HOPG analyzer to lower background scattering (providing approximately 1 meV energy resolution). A thick block of HOPG was used to filter out the $\lambda/2$ component from the incident beam.  The data between $2 <T<300$ K were collected using an 
{\it orange} cryostat and a high temperature furnace was used for the 
measurements between $300<T<420 $ K. All  the neutron diffraction data were analyzed using 
Rietveld refinement program Fullprof  suite\cite{Fullprof}. 

The temperature and magnetic field dependence of the magnetization were carried out in a 
superconducting quantum  interference device (Quantum Design MPMS-7S, SQUID) magnetometer. 

\section{RESULTS AND DISCUSSION}
\subsection{ A. Crystalline structure }
\begin{figure} \centering \includegraphics [width = 1\linewidth] {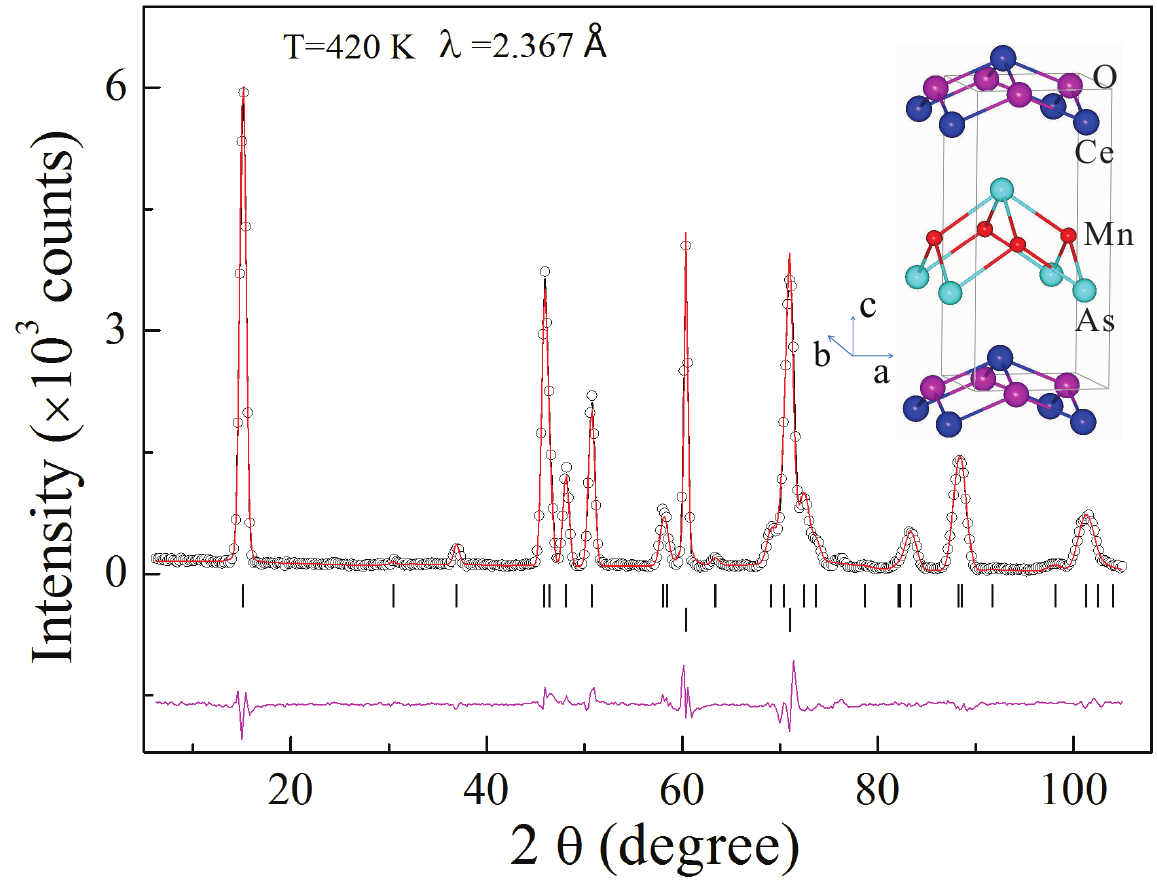}
\caption{(color online)  Rietveld refinement fit to neutron diffraction pattern 
at 420 K and a graphic 
representation of the crystal structure
for CeMnAsO using the best fit parameters listed in Table I. The observed data 
and the fit are indicated by the open circles and 
solid lines, respectively. The difference curve is shown at the bottom. The 
vertical bars mark the positions of Bragg reflections for the phases of CeMnAsO 
(up) and Al sample holder (below).}
\label{fig:Rietveld} 
\end{figure}

Neutron powder diffraction pattern at 420 K is shown in Fig.\ \ref{fig:Rietveld} indicating the 
purity of the material with no indication of a secondary phase in CeMnAsO, 
consistent with our x-ray diffraction measurement at room temperature (not shown 
here). Rietveld analysis confirms the tetragonal ZrCuSiAs-type structure with 
space group $P4/nmm$, as illustrated in the inset of Fig.\ \ref{fig:Rietveld}. 
Similar to 
the tetragonal structure in \textit{R}FeAsO (\textit{R} is rare earth 
element)\cite{Zhang2013} or LaMnAsO \cite{Hanna2013}, the structure of CeMnAsO 
consists of MnAs and CeO layers where the Mn$^{2+}$ ions form a square lattice.  
Our neutron diffraction results show no change in the crystal structure of this 
compound down to 2 K.  The refined atomic positions, lattice constants, and  
volume of CeMnAsO at 420 K and ground temperature 2 K are summarized in Table \  
\ref {tab:lattice}.

\begin{figure} \centering \includegraphics [width = 1\linewidth] {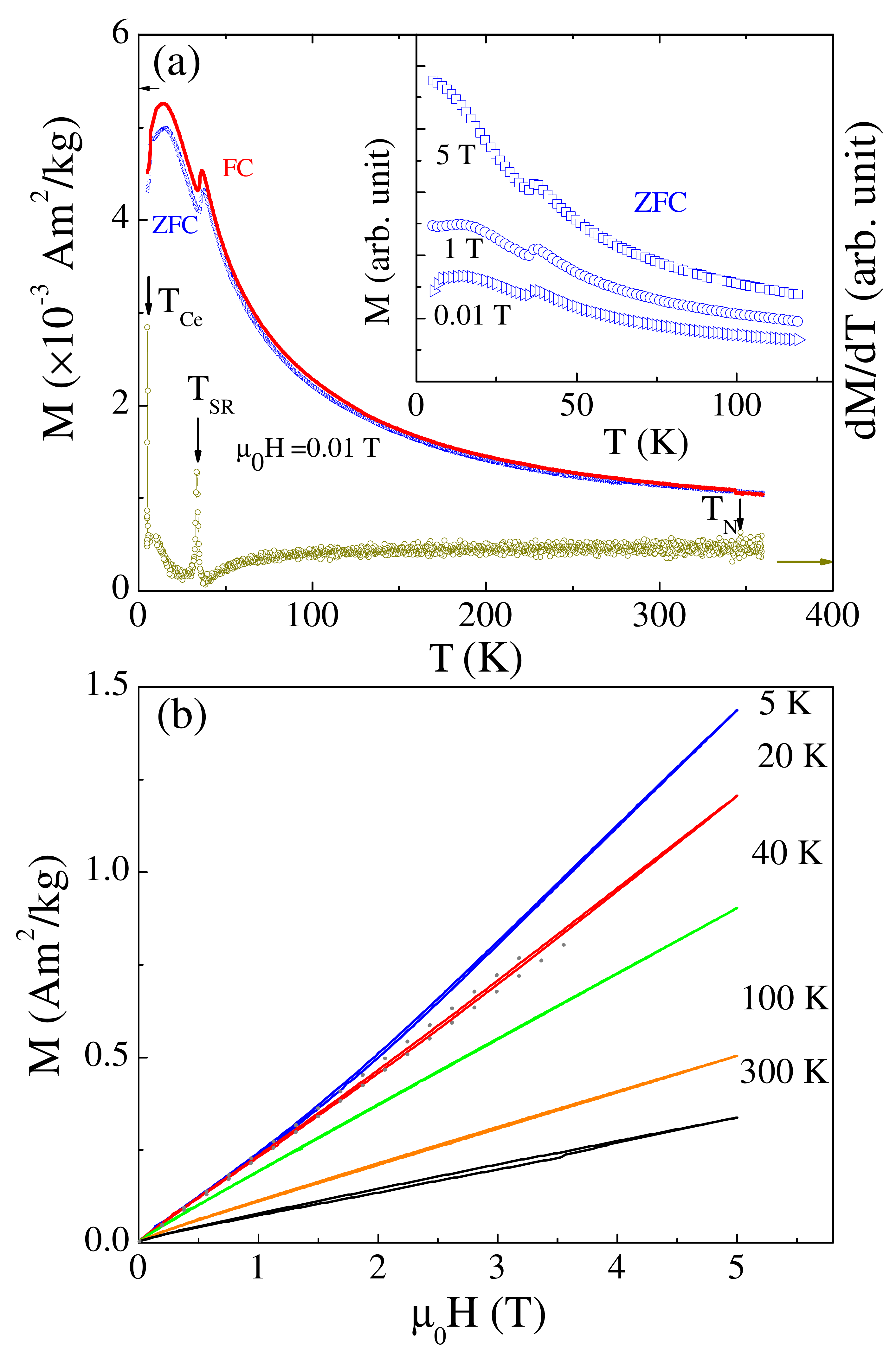}
\caption{(color online)  (a) Temperature dependence of the zero-field-cooling 
(ZFC) and FC magnetization in a low field of 0.01 T for CeMnAsO. The curve at 
the bottom shows the first derivative of FC curve with no indication of $T_{\rm 
N} = 345 K$.  The inset shows the temperature dependence of the ZFC  
magnetization at 0.01, 1 and 5 T. (b) Field dependence of magnetization at 
different temperatures in CeMnAsO.}
\label{fig:Mag} 
\end{figure}
\subsection{ B. Multiple magnetic transitions and 
field-induced metamagnetic transition revealed in magnetization measurements}
\begin{figure} \centering \includegraphics [width = 1\linewidth] {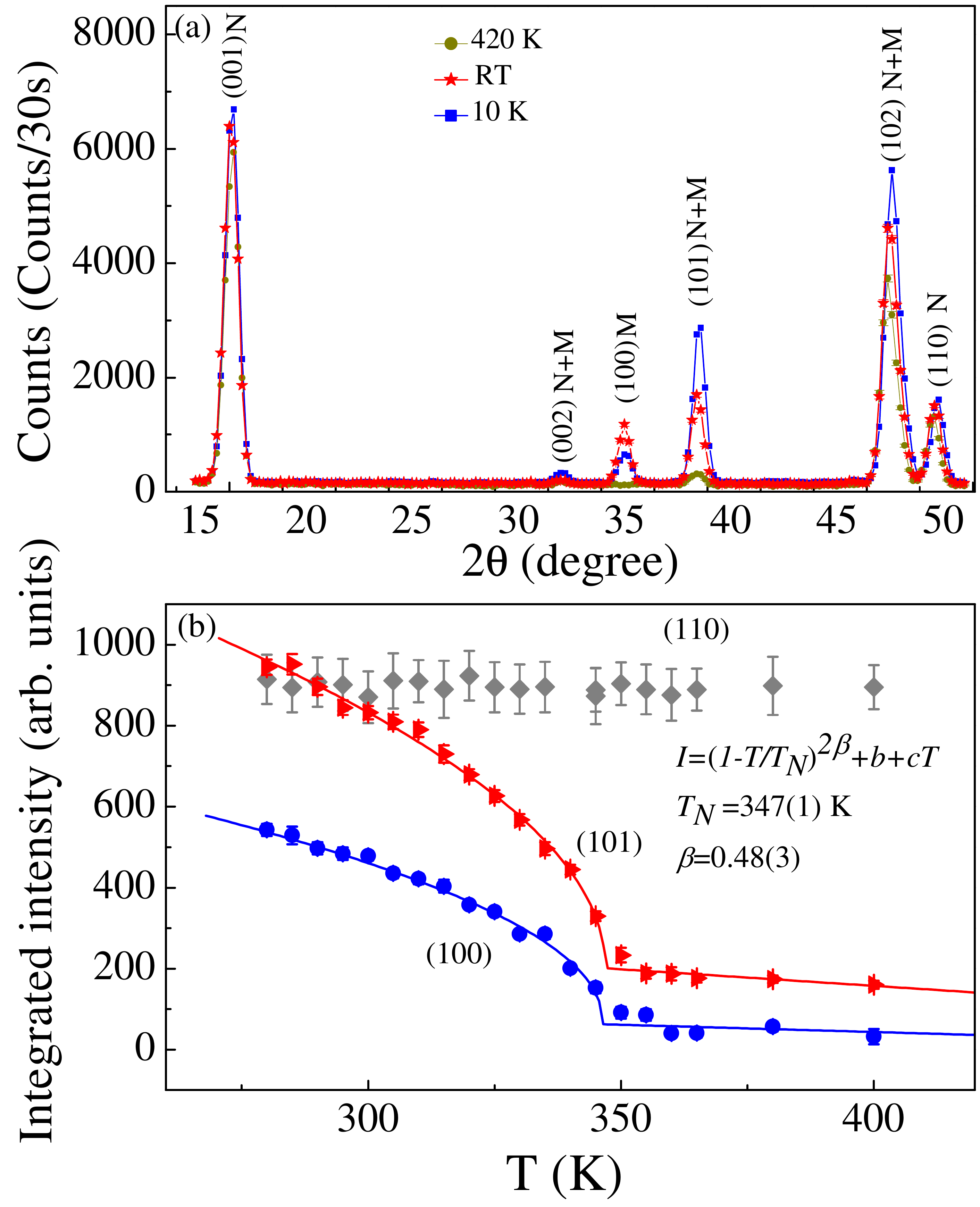}
\caption{(color online) (a) Comparison of the neutron diffraction patterns at 
420 K, RT and 10 K. (b) Integrated intensity of (110), (100) 
and (101) reflections as 
a function of temperature with an fit to a power law with relevant parameters 
as shown in the figure.}
\label{fig:OrderPara1} 
\end{figure}

The temperature dependence of the zero-field-cooled (ZFC) and field-cooled (FC)  
magnetization in Fig.\ \ref{fig:Mag}(a) shows a clear magnetic transition at 35 K emphasized 
by a single peak in the first derivative of magnetization with no indication of 
additional anomaly up to 370 K. The anomaly at 35 K has been previously 
attributed to  a spin reorientation (SR) transition of Mn.\cite{TSUKAMOTO2011} 
Our susceptibility measurements show that this magnetic transition is not 
shifted by external magnetic fields up to 5 T in accordance with typical 
behavior of a spin reorientation transition (the transition temperature at 35 K 
is labeled $T_{SR}$ hereafter). The thermal hysteresis of the magnetization  
below T$_{SR}$ is indicative of the first-order nature of the transition.  Below 
7 K, both ZFC and FC magnetization decrease implying the emergence of another 
magnetic transition. We point out that such anomalous decrease below 7 K was  
not observed by Tsukamoto \textit{et al.} presumably because of the influence of a 
secondary phase as mentioned in Ref. 12. Interestingly,  we do not observe a 
clear anomaly in the magnetization in the temperature range  $35 - 370$ K potentially identifying $T_{\rm N}$. 
However, the neutron measurements of the (100) and (101) magnetic Bragg peaks 
shown in Fig.\ \ref{fig:OrderPara1} (b) exhibit a sharp increase in the 
integrated intensity below $\approx 345$ 
K, which we identify as the AFM transition temperature $T_{\rm N}$ of the Mn 
sublattice. We note that a weak and broader (100) Bragg peak persists above 
$T_{\rm N}$ with a linewidth that increases with  temperature indicating the 
presence of short-range ordered Mn spins above $T_{\rm N}$, as shown in the 
inset of Fig. 4(a). This suggests the 
existence of strong spin fluctuations above $T_{\rm N}$ that tend to wash out 
any anomaly in the susceptibility at $T_{\rm N}$ even in its first derivative 
with respect to temperature. The absence or weak peak in the derivative of the 
susceptibility is characteristic of the two-dimensional nature of the Mn 
magnetic system with a strong inplane coupling that gives rise to short range 
fluctuating magnetic order, as has been found in other systems 
\cite{Vaknin1989,Vaknin1990}.  This is consistent with the 
overall behavior of the order parameters as a function of temperature as 
expressed in the (100) and the (101) magnetic refelctions shown in Fig.\ 
\ref{fig:OrderPara1} (b).  The intensity of both peaks is modeled by a power 
law 
\begin{equation}
I(T) = a(1-T/T_N)^{2\beta} +b+cT
\end{equation}   
where $a$ is an intensity scale factor (at $T=0$) and $b$ and $c$ account for 
background and signal temperature dependent above $T_{\rm N}$.  Our fit to both 
peaks yields $T_{\rm N} =347\pm1$ K and 
$\beta = 0.47\pm0.03$.  The overall temperature dependence with a relatively large $\beta$ (compared to $\beta = 0.125$ for the 2D Ising model or $\beta \approx 0.36$ for the 3D Heisenberg model) has been explained for similar 2D system with inplane exchange coupling $J_1$ that is much larger than interlayer one $J_c$, $J_c/J_1 << 1$ \cite{Singh1990}.
\begin{table}
\caption{Refined atomic positions, lattice constants, lattice volume $V$ at 
$T = 420$ and  2 K for CeMnAsO space group $P4/nmm$. Ce: 
2\textit{c}($\frac{1}{4}$, 
$\frac{1}{4}$, z); Mn: 2\textit{b} ($\frac{3}{4}$, 
$\frac{1}{4}$, $\frac{1}{2}$); As: 2\textit{c} ($\frac{1}{4}$, 
$\frac{1}{4}$, z); O: 2\textit{a} ($\frac{3}{4}$, 
$\frac{1}{4}$, 0). } 
 \label{tab:lattice}
%\begin{ruledtabular}
\begin{tabular} {llllll}
 \hline\hline
T & Atom& Atomic position & \textit{a}(\AA{})& \textit{c}(\AA{}) & 
 $V$(\AA{}$^{3}$) \\
\hline
420 K&   Ce  &   z= 0.1263(6)& 4.1032(2) & 9.0038(3) &151.594(8)\\
 &   As &    z= 0.6696(4)     &   &  &  \\
 2 K&   Ce  &   z= 0.1287(4)& 4.0837(3) & 8.9567(4) &149.370(6)\\
 &   As &    z= 0.6720(5)     &   &  &  \\
 \hline\hline
\end{tabular}
%\end{ruledtabular}
\end{table}

Figure\ \ref{fig:Mag} (b) shows magnetic field dependence of the magnetization at different 
temperatures. Whereas the magnetization at room-temperature (RT) shows a very 
weak hysteresis, typical of AFM behavior, i.e., linear $M$ versus $H$ curve is 
observed in $T_{\rm SR}<T<T_{\rm N}$.  Below $T_{\rm SR}$, the magnetization 
first rises linearly at low fields indicative of an AFM behavior, with a weak 
jump in the slope at approximately 1.5 T, indicating a possible field-induced 
meta-magnetic transition. However, the magnetization is far from  saturation 
even at 5 T, indicating that the magnetic structure under magnetic field is 
mostly preserved, and that the external magnetic field  induces a 
transformation to weakly canted AFM structures below $T_{SR}$. 

\begin{figure} \centering \includegraphics [width = 1\linewidth] {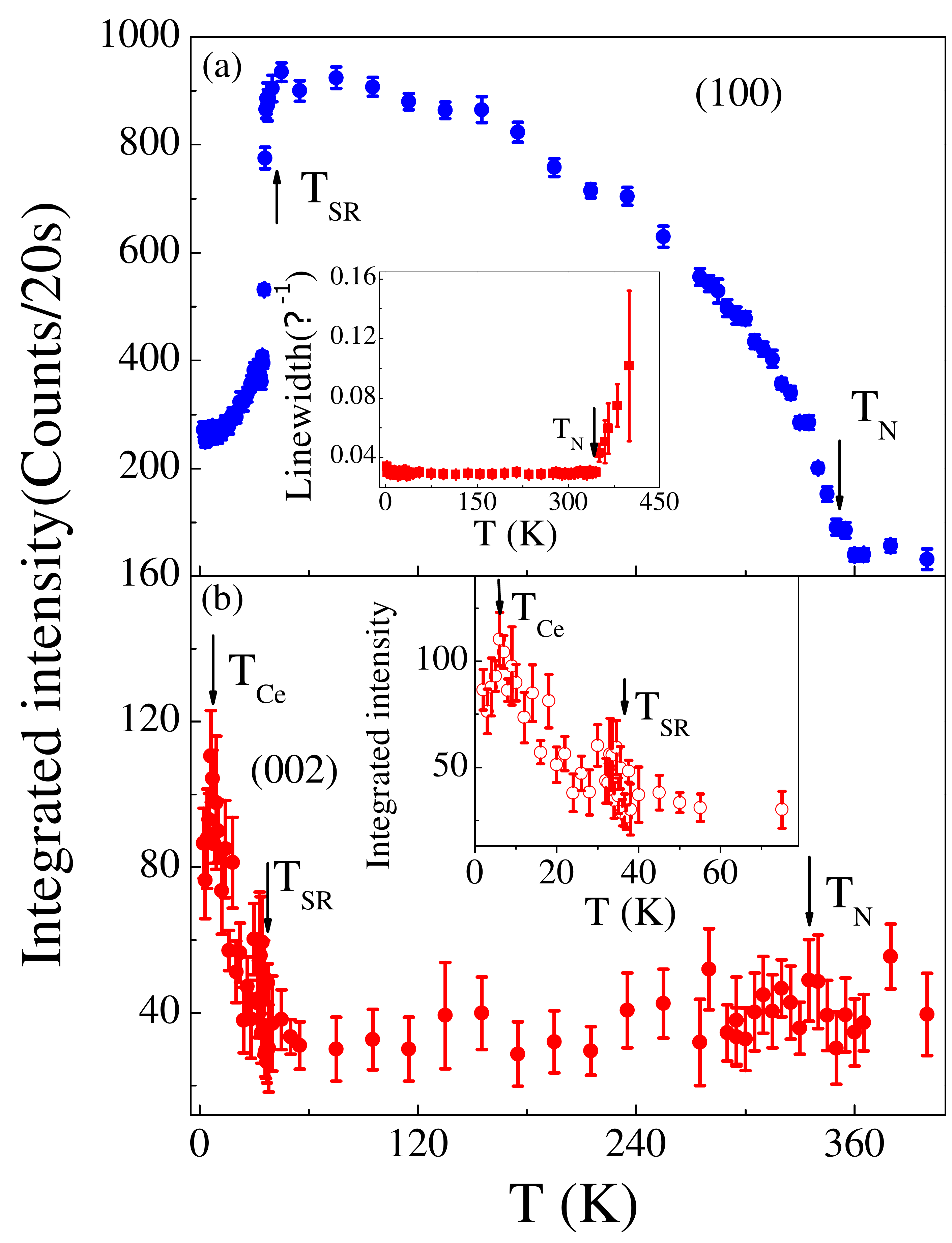}
\caption{(color online) (a) Temperature dependence of the 
integrated intensity of (100)  peak. The inset shows the temperature dependence 
of the linewidth of (100) peak. (b) Temperature dependence of the integrated 
intensity of (002) peak. The inset shows a zoomed-in view at low temperatures 
below 80 K with a peak at $T = 7$ K that we attribute to additional magnetic 
transition in the Ce sublattice.  }
\label{fig:OrderPara2} 
\end{figure}

\subsection{ C. Evidence of Mn spin reorientation and 
long-range Ce spin orderings determined by NPD measurements
}
A comparison of the neutron diffraction patterns at different temperature  
windows $T>T_{N}$ (420 K), $T_{\rm SR}<T<T_{\rm N}$ (RT), $7 K<T<T_{SR}$  (10K) 
is shown in Fig.\ \ref{fig:OrderPara1} (a) (nuclear and magnetic Bragg reflections are labeled N and 
M, respectively).  At RT, whereas the (100) Bragg peak is purely magnetic, the 
(101) and (102) peaks have nuclear and magnetic contributions due to the Mn 
ordering. At 10 K (below $T_{\rm SR}$), the intensity of the (100) peak decreases dramatically whereas the intensities of the (101) and (102) increase, evidence for a change in 
magnetic structure. The magnetic contribution to the (002) nuclear peak  below 
$T_{N}$ is negligible, however it increases below $T_{\rm SR}$ and peaks at 
$\approx 7$ K (see Fig.\ \ref{fig:OrderPara2} (b)).  As discussed below, the 
intensities of the (100) and (002) peaks reflect the order parameters of Mn and Ce 
moments, respectively.  All the magnetic reflections can be indexed on the high 
temperature nuclear (chemical) unit cell with a magnetic  propagation vector 
$k=(0,0,0)$.  SARAH representational analysis program \cite{SARAH} is used to 
derive the symmetry allowed magnetic structures. The decomposition of the 
magnetic representation ($\Gamma_{ Mag}$) into the irreducible representations 
is $\Gamma^{1}_{3}+\Gamma^{1}_{6}+\Gamma^{2}_{9}+\Gamma^{2}_{10}$ and 
$\Gamma^{1}_{2}+\Gamma^{1}_{3}+\Gamma^{2}_{9}+\Gamma^{2}_{10}$ for Mn sites  and 
Ce sites, respectively.  The symmetry allowed basis vectors are summarized in 
Table \  \ref {tab:MagMoment}. There are two FM ($\Gamma^{1}_{3}$ and 
$\Gamma^{2}_{9}$) and three AFM ($\Gamma^{1}_{2}$, $\Gamma^{1}_{6}$ and 
$\Gamma^{2}_{10}$) solutions.  But the two FM solutions can be discarded at all 
temperatures as there is no FM contribution to the nuclear Bragg reflections in 
our neutron diffraction patterns consistent with the magnetization measurements 
below $T_{\rm N}$. Thus, only the three AFM solutions are considered for the 
data refinement to obtain the magnetic structures at different temperature 
windows.

\begin{figure} \centering \includegraphics [width = 1\linewidth] {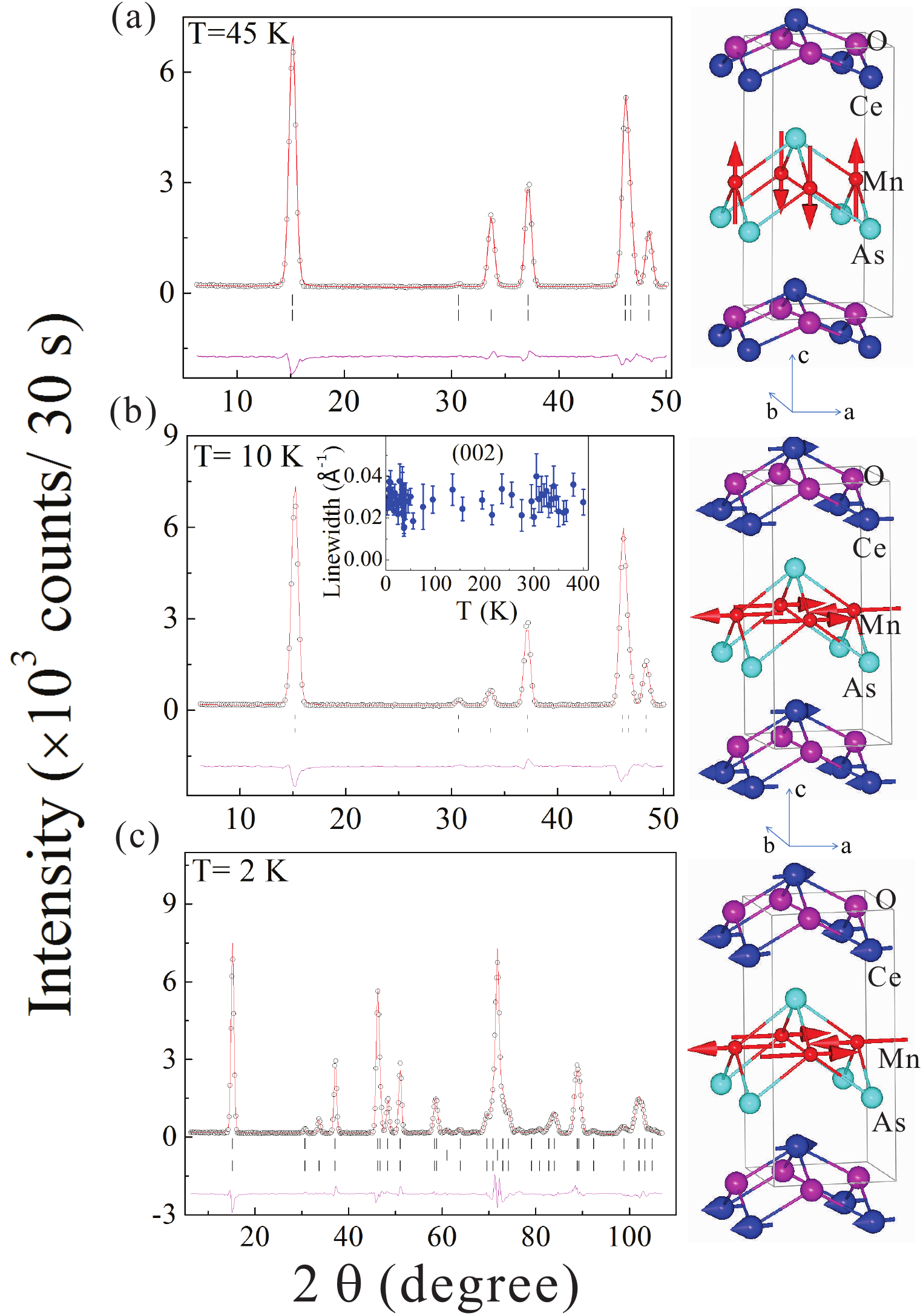}
\caption{(color online)  Rietveld refinement fits to neutron diffraction 
patterns and the graphic representations of the determined magnetic structures 
of CeMnAsO at (a)  45 K,(b) 10 K and (c) 2 K. The observed data and the fit are 
indicated by the open circles and solid 
lines, respectively. The difference curve is shown at the bottom. The vertical 
bars mark the 
positions of Bragg reflections for the nuclear phase (up) and magnetic phase 
(down) in CeMnAsO. The middle vertical bars in (c) mark the positions of the 
phase of Al sample holder. The inset of (b) shows the temperature dependence of 
the linewidth of (002) peak.}
\label{fig:MagneticStructures} 
\end{figure}  

Rietveld refinement fits to neutron diffraction patterns and the graphic 
representation of the determined magnetic structures of CeMnAsO at different 
temperatures are shown in Fig.\ \ref{fig:MagneticStructures}.  In $T_{\rm SR}<T<T_{\rm N}$, there is no 
evidence for Ce moment ordering  and the neutron diffraction pattern is best 
fitted using $\Gamma^{1}_{6}$ model, i.e., the Mn spins are antiparallel at Mn1 
and Mn2 sites, forming a nearest-neighbor antiferromagnetic alignment in 
\textit{ab} plane and the planes are stacked ferromagnetically along 
\textit{c}-axis, i.e., C-type AFM 
order with the Mn moments along \textit{c}-axis, as shown in Fig.\ 
\ref{fig:MagneticStructures} (a) at 45 K. As the temperature decreases to 
$T_{\rm SR}$, the 
magnetic structure is preserved and the Mn magnetic moment gradually increases 
with an average moment 2.29(3) $\mu_{B}$ at RT and  2.78(2) $\mu_{B}$ at 45 K.
   
\begin{table}
\caption{The symmetry-allowed basis vectors [m$_{x}$,m$_{y}$,m$_{z}$] for the 
space group $P4/nmm$ with \textbf{k}=(0,0,0) in CeMnAsO. Mn1: (0.75, 0.25, 
0.5), Mn2:(0.25, 0.75, 0.5), Ce1:(0.25, 0.25, 0.126) and Ce2:(0.75, 0.75, 
0.874). } 
 \label{tab:MagMoment}
%\begin{ruledtabular}
\begin{tabular} {llllll}
 \hline\hline
Atom& $\Gamma^{1}_{2}$  & $\Gamma^{1}_{3}$ & $\Gamma^{1}_{6}$ & 
$\Gamma^{2}_{9}$&$\Gamma^{2}_{10}$\\
  &&&& \\
\hline
Mn1 &     &   [0 0 m$_{z}$] & [0 0 m$_{z}$]  & [m$_{x}$ m$_{y}$ 0] &[m$_{x}$ 
m$_{y}$ 0]\\
Mn2&    &    [0 0 m$_{z}$]  &    [0 0 -m$_{z}$]& [m$_{x}$ m$_{y}$ 0] &[-m$_{x}$ 
-m$_{y}$ 0] \\
Ce1& [0 0 m$_{z}$] &   [0 0 m$_{z}$]  &    & [m$_{x}$ m$_{y}$ 0]&[m$_{x}$ 
m$_{y}$ 0]\\
Ce2&[0 0 -m$_{z}$]  & [0 0 m$_{z}$]      &   & [m$_{x}$ m$_{y}$ 0]&[-m$_{x}$ 
-m$_{y}$ 0] \\
 \hline\hline
\end{tabular}
%\end{ruledtabular}
\end{table}    
   
In the temperature range 7 K $ <T<T_{SR}$, the refinement of the neutron 
diffraction patterns are not satisfactory with the assumption of Mn ordering 
only and the ordering of  Ce magnetic moments is required to obtain good 
agreement with the data. Trial refinements assuming a linear combination 
of the $\Gamma^{1}_{6}$ of Mn sites and $\Gamma^{1}_{2}$ of Ce sites, $\Gamma^{2}_{10}$ 
of Mn sites and $\Gamma^{1}_{2}$ of Ce sites, or $\Gamma^{1}_{6}$ of Mn sites 
and $\Gamma^{2}_{10}$ of Ce sites do not fit the data well. A satisfactory fit 
to the diffraction patterns below $T_{\rm SR}$ is obtained by using 
$\Gamma^{2}_{10}$, i.e., antiparallel Ce spins at Ce1 and Ce2 sites and 
antiparallel Mn spins at Mn1 and Mn2 sites with ordered Mn and Ce moments in the 
\textit{ab} plane.  Thus, Mn maintain the C-type magnetic structure
but the ordered Mn magnetic moments reorient to the \textit{ab} plane, and simultaneously 
the Ce spins align antiferromagnetically along \textit{c},  
similar to the magnetic structure in  
PrMnSbO\cite{Kimber2010} and NdMnAsO\cite{Emery2011} below its SR transition. 
It is impossible to determine the absolute angle between Mn (or Ce) moments with respect to 
\textit{a} axis in \textit{ab} plane due to the tetragonal nature of the 
system, nevertheless we show both Mn and Ce moments along \textit{a} axis in 
Fig.\ \ref{fig:MagneticStructures}(b).  Note that there is no significant broadening of the (002) peak 
below/above $T_{\rm SR}$, as shown in the temperature dependence of its 
linewidth (see the inset of Fig.\ \ref{fig:MagneticStructures} (b)) indicating 
long-range ordered Ce below $T<T_{SR}$.

\begin{figure} \centering \includegraphics [width = 1\linewidth] {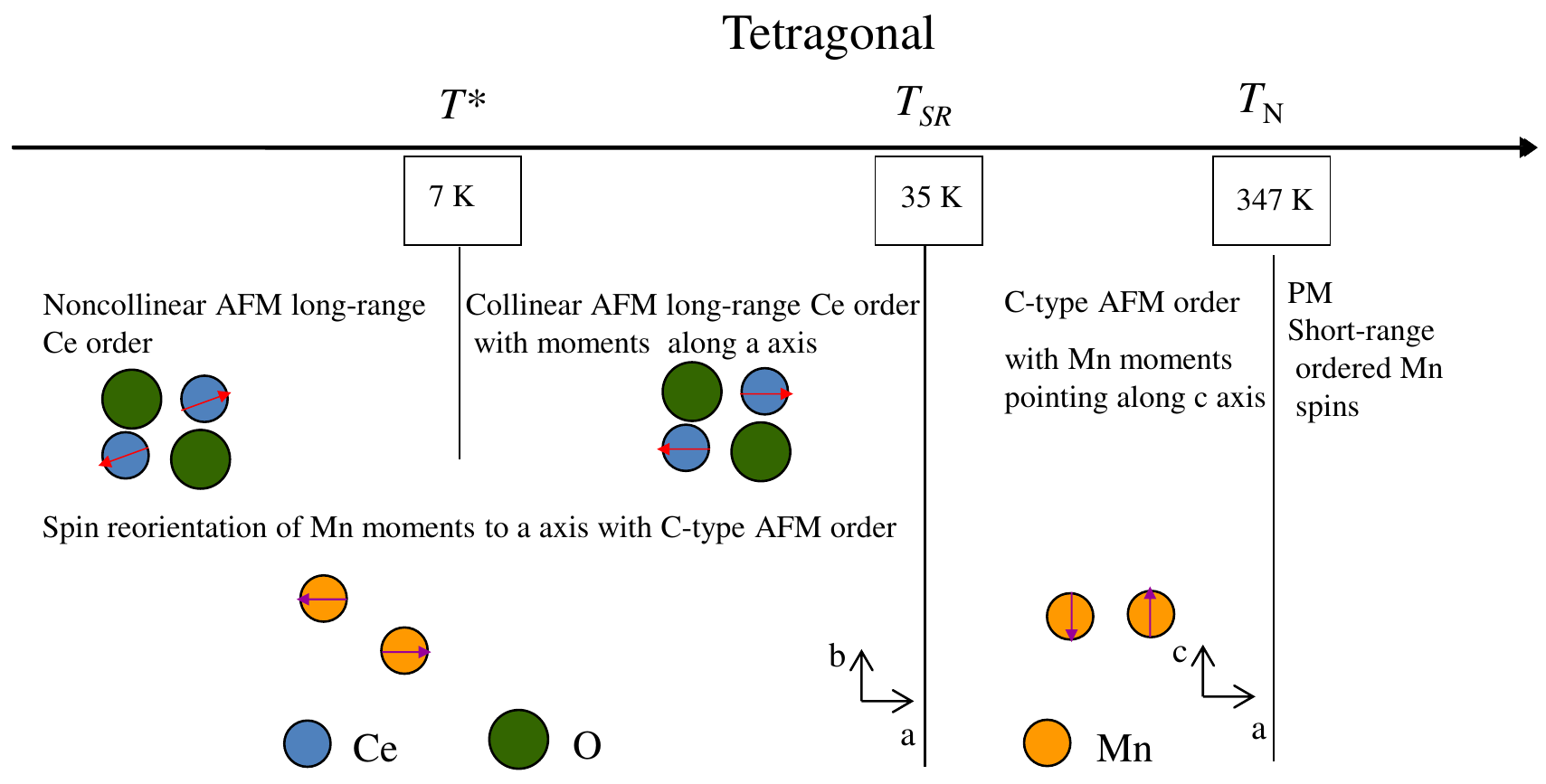}
\caption{(color online) Schematic illustration of the proposed magnetic 
structures for Ce and Mn sublattices in the CeMnAsO. Note that the different 
\textit{ac }and \textit{ab} planes are used to illustrate the magnetic 
structures above and below $T_{SR}$, respectively. } 
\label{fig:PhaseDiagram} 
\end{figure}

Below $T_{\rm SR}$,  whereas there is no anomaly at 7 K in the intensity of the 
(100) 
reflection (see Fig.\ \ref{fig:OrderPara2} (a)), the intensity of the (002) 
reflection increases slightly 
peaking at $\sim 7$ K (see Fig.\ \ref{fig:OrderPara2} (b)), consistent with the 
peak in the 
magnetization shown in 
Fig.\ \ref{fig:Mag}(a), which confirms there is 
another magnetic transition in the Ce sublattice. 
Furthermore, the good refinement of the neutron diffraction pattern at 2 K is 
still obtained by using $\Gamma^{2}_{10}$ model within uncertainties. Since the 
refinement using  $\Gamma^{2}_{10}$ model of Ce1 and Ce2 spins requires 
antiparallel and confined spins to the \textit{ab} plane, it is very likely that 
the transition observed at 7 K is due to a finite angle between Ce and Mn spins 
with 
respect to the Mn spins, forming a noncollinear magnetic structure between Ce 
and Mn moments as theoretically predicted\cite{Lee2012}.  However, we emphasize 
that our powder data is not sufficiently sensitive to determine the accurate 
relative angle between Ce and Mn moments. The quality of the refinement to the 2 
K data has a tendency to be slightly improved when this angle is increased 
from 0$^{\rm o}$ to $\sim20^{\rm o}$. The best refinement results using a 
20$^{\rm o}$ angle are shown in Fig.\ \ref{fig:MagneticStructures}(c).  
The average ordered moments at 2K for  Mn is  3.32(4) $\mu_{B}$ and for Ce is in the range of 
$0.75(3) - 0.81(4)\mu_{B}$ (depending on the relative angle between Ce and Mn 
moments). The reduced Mn ordered moment (from that of $S=5/2$; $\sim 5$ $ 
\mu_B$ for an ideal localized moment) in CeMnAsO is likely due to the 
spin-dependent hybridization between the Mn  3\textit{d} and As 4\textit{p} 
orbitals as in BaMnAsF\cite{Saparov2013} and BaMn$_{2}$As$_{2}$\cite{An2009} 
that all share similar MnAs layer.  Such hybridization was also noted for iron based pncitides such as SrFe$_2$As$_2$\cite{Lee2010}.  It is worthwhile noting that the  (002) peak 
intensity and the peak in the susceptibility at 7 K in CeMnAsO are different 
from the observations in the iso-structural NdMnAsO in which both the intensity 
of the magnetic peak and  the susceptibility saturate below 4 K with no 
indication of a magnetic transition in the Nd sublattice \cite{Emery2011}. Our 
proposed schematic illustration  of the magnetic transitions in CeMnAsO is 
summarized in Fig.\ \ref{fig:PhaseDiagram}.

\subsection{  D. Magnetic interactions and absence of T-O structural 
transitions above and below $T_{\rm N}$}
The ordered Mn moments  3.32(4) $\mu_{B}$ at 2 K compared to the 5 $\mu_B$  
expected from a localized moment indicate that in the spectrum of itinerant {\it 
versus} local-moment AFM, CeMnAsO tends to be the latter (i.e., local-moment 
AFM)\cite{TSUKAMOTO2011}. This is different from the itinerant CeFeAsO with a 
much lower ordered Fe moment of $\sim0.9 \mu_{B}$ compared to the 4 $\mu_B$ 
expected from a localized moment\cite{Zhang2013}.  The inplane 
checker-board-like AFM structure of the C-type order in CeMnAsO in $T<T_{\rm N}$ 
suggests that the NN interaction $J_{1}$ is dominant whereas inplane  
next-nearest-neighbor (NNN) interaction $J_{2}$ is very weak or negligible. 
Thus, in the context of $J_{1}-J_{2}-J_{c}$ model  \cite{Johnston2011}, we 
conclude that $J_{1}>0$, $J_{2}<J_{1}/2$ with 
negligible spin frustration. This is in sharp contrast to CeFeAsO for which the 
effective NNN interaction $J_{2}>J_{1}/2$\cite{Calderon} is necessary to 
stabilize the 
stripe-like AFM ordering with the ordered moment in the \textit{ab} plane. Note 
that the preferred orientation of Mn is along the \textit{c} axis in CeMnAsO, in 
contrast to the preferred orientation of Fe in \textit{ab} plane in CeFeAsO. 
  
Another significant difference between CeMnAsO and CeFeAsO is that the T-O 
structural transition observed above $T_{\rm N}$ in CeFeAsO is absent in 
CeMnAsO. It is generally accepted that the T-O 
structural transition in CeFeAsO and other ``1111" Fe-based pnictides
has a magnetoelastic origin\cite{Singh2009} due to nematic 
fluctuations\cite{Fernandes2014,Zhang2014}. The stripe-like AFM structure in 
``1111" Fe-based pnictides can be separated into two N\'{e}el sublattices each 
defined by NNN Fe spins in the basal plane.\cite{Fernanders2010} Due to 
$J_{2}>J_{1}/2$, there is a strong frustration between the two sublattices and the orthorhombic
distortion reduces the frustration and lowers the total energy (magnetic and elastic energy)\cite{Singh2009,Fernandes2014, Fernanders2010}. The absence of such magnetic 
frustration may explain the absence of a T-O  structural transition in CeMnAsO .  This 
is further consistent with the  absence of the T-O transition in the isostructural 
BaMnAsF\cite{Saparov2013}, and in the Mn-based``122'' systems 
BaMn$_{2}$As$_{2}$\cite{Singh2009}. The AFM structure in both systems is G-type 
with the NN Mn spins antiparallel along all the directions and moments 
along the $c$-axis. We note that the T-O  transition at $ \approx 35 $ K found 
in 
PrMnSbO below its $T_{\rm N} =230$  K is likely driven by  local 
\textit{f}-electrons in  Pr$^{3+}$, unlike the T-O transition in the ``1111" 
Fe-based pnictides driven by the transition metal. Different from PrMnSbO, no 
any structural transition below $T_{\rm N}$ is oberved in CeMnAsO, the 
mechanism of
which deserves further investigations.

As compared to BaMn$_{2}$As$_{2}$ with  antiparallel Mn spins along 
\textit{c}-axis, the parallel Mn spins along \textit{c}-axis
in CeMnAsO suggests that the interlayered magnetic interaction $J_{c}>0$ in 
BaMn$_{2}$As$_{2}$ but $J_{c}<0$ in CeMnAsO. The N\'{e}el temperature of 625 K 
in BaMn$_{2}$As$_{2}$ with three-dimensional magnetism  is much higher than 347 
K in CeMnAsO.  Assuming the inplane exchange coupling $J_1$ is of the same order of magnitude   this implies a much weaker magnetic interlayer interaction 
$J_c$ in CeMnAsO, however with strong 2D correlations (fluctuated) as discussed above. 
The much weaker interlayer magnetic interaction due to the longer distance 
between adjacent 
MnAs layers leads to a quasi-two-dimensional AFM character in CeMnAsO 
similar to other ``1111" systems\cite{Memet2013}. 

For the doped Fe-based 
superconductors, it is commonly observed that the emergence of the SC is 
accompanied with the suppression of both the structural and magnetic transitions 
with $J_{2}>J_{1}/2$. However, for CeMnAsO, there is no evidence for 
 T-O structural transition and $J_{2}<J_{1}/2$. Further, CeMnAsO is a 
local 
moment antiferrromagnet in contrast to the itinerant antiferromagnet in 
Fe-based 
superconductors. This indicates that Mn at the transition metal site may 
prevent the emergence of 
superconductivity in Mn-doped ''1111" and ''122" Fe-based pnictides, which is 
supported by experimental evidence that the substitution of Mn for Fe 
in Ba(Fe$_{1-x}$Mn$_{x}$)$_{2}$Mn$_{2 }$ does not induce SC
\cite{Thaler2011}. 

\subsection{E. Ce-Mn coupling}
The Mn SR transition  is not observed in 
LaMnAsO \cite{Emery2011}, BaMnAsF\cite{Saparov2013} or 
BaMn$_{2}$As$_{2}$ \cite{Singh2009} where there is no magnetic rare earth ion 
but is found in CeMnAsO and NdMnAsO \cite{Emery2011}, which indicates the SR 
transition in CeMnAsO is driven by the coupling between rare earth Ce and Mn. 
The Mn$^{2+}$ moment, 
commonly displays very weak single-ion anisotropy as expected for the $L=0$ of 
Mn$^{2+}$\cite{Toft-Petersen2012}, favors orientation along the \textit{c}-axis. As soon as Ce$^{3+}$ 
spins ($S= 1/2$ and $L= 3$) order below 35 K, the Ce-Mn coupling  exerts an 
effective field that induces a flop of  Mn$^{2+}$ spins to the basal plane.  
We point out that there is also a spin reorientation  in CeFeAsO 
\cite{Zhang2013} and 
CeCrAsO \cite{TSUKAMOTO2011} but not 
in CeCoAsO\cite{Sarkar2010} or CeNiAsO \cite{Luo2011}. In CeFeAsO, the 
stripe-like Fe$^{2+}$ spins rotate uniformly and gradually in the \textit{ab} 
plane below $ \approx14$ K and the SR of Cr occurs at $ \approx 36$ K. Recently 
we performed neutron diffraction measurements on CeCoAsO and confirmed the FM 
behavior of Co without evidence for a Co SR transition or Ce ordering  in 
agreement with  previously reported studies\cite{Sarkar2010}. SR of transition 
metal ions Mn, Fe, Cr, has also been observed in other systems due to the  
coupling between magnetic rare earth \textit{f} and transition metal \textit{d} 
moments, such as in \textit{R}FeO$_{3}$ (\textit{R}=Ce and Nd) \cite{RFeO}, 
\textit{R}Fe$_{2}$ (\textit{R}= Ce and Nd) \cite{Belov1976}, 
\textit{R}$_{2}$Fe$_{14}$B (\textit{R}=Nd and Er)\cite{Guslienko1995}, hexagonal 
HoMnO$_{3}$ \cite{Vajk2005}, \textit{R}CrO$_{3}$ (\textit{R}=Ce, Nd and 
Sm)\cite{Cao2014}.

\section{CONCLUSION}
In summary, we report on the structure and magnetic properties in CeMnAsO. 
Whereas no structural transition is observed above and below the N\'{e}el 
temperature in tetragonal CeMnAsO, it exhibits 
a set of complex magnetic transitions. We find 
two-dimensional short-range ordered Mn (most likely dynamic in nature, i.e., 
spin fluctuations) 
above $T_{\rm N}=347(1)$ K.  Below $T_{\rm N}$, the Mn spins order in  a 
C-type AFM structure with moments pointing along the \textit{c}-axis. A  spin 
reorientation of the Mn moments 
from the $c$-axis to the \textit{ab} plane while keeping the C-type order occurs 
below $T_{\rm SR}=35$ K, which is induced by  long-range ordering of the Ce via 
Ce-Mn coupling. Below 7 K, the collinear magnetic structure transforms to a 
noncollinear one with an angle between the Ce and Mn moments. 
A possible field-induced metamagnetic transition is observed below $T_{\rm 
SR}$ in magnetization measurements. 
The local-moment 
antiferromagnetism with dominant NN  interaction and negligible NNN interaction 
with $J_{2}<J_{1}/2$ 
in CeMnAsO contrasts with the itinerant antiferromagnetism in CeFeAsO with 
$J_{2}>J_{1}/2$.  We also point out that the spin reorientation 
transition is common not only to Mn, but also to Fe or Cr ions in the 
oxypnictides and other oxides or intermetallics induced by strong 
coupling between rare-earth \textit{R} and the transition metal ions.

\section{Acknowledgments}
Research at Ames Laboratory is supported by the US Department of Energy, Office 
of Basic Energy Sciences, Division of Materials Sciences and Engineering under 
Contract No. DE-AC02-07CH11358. Use of the high flux isotope reactor at the Oak 
Ridge National Laboratory, was supported by the US Department of Energy, Office 
of Basic Energy Sciences, Scientific User Facilities Division.

\end{document}